\documentstyle[twoside,fleqn,espcrc2]{article}

\begin{document}

% put your own definitions here:
\def\PL #1 #2 #3 {{\rm Phys. Lett.} {#1} (#3) #2}
\def\NP #1 #2 #3 {{\rm Nucl. Phys.} {#1} (#3) #2}
\def\ZP #1 #2 #3 {{\rm Z. Phys.} {#1} (#3) #2}
\def\PRL #1 #2 #3 {{\rm Phys. Rev. Lett.} { #1} (#3) #2}
\def\PR #1 #2 #3 {{\rm Phys. Rev.} {#1} (#3) #2}
\def\MPL #1 #2 #3 {{\rm Mod. Phys. Lett.} {#1} (#3) #2}
\def\RMP #1 #2 #3 {{\rm Rev.~Mod. Phys.} {#1} (#3) #2}
\def\ifm{\ifmmode}
\def\al{\ifm \alpha \else $\alpha$\fi}
\def\als{\ifm \alpha_s \else $\alpha_s$\fi}
\def\go{\ifm \rightarrow \else $\rightarrow$\fi}
\def\shat{\ifm \hat{s} \else $\hat{s}$\fi}
\def\shh{\ifm \hat{\hat{s}} \else $\hat{\hat{s}}$\fi}
\def\that{\ifm \hat{t} \else $\hat{t}$\fi}
\def\gam{\ifm \gamma \else $\gamma$\fi}
\def\thep{\ifm \theta ' \else $\theta '$\fi}
\def\pt{\ifm p_{\rm T} \   \else $p_{\rm T}$\ \fi}
\def\ptm{\ifm p_{\rm T\min} \  \else $p_{\rm T\min}$\ \fi}
\def\qb{\ifm \bar{q} \else $\bar{q}$\fi}
\def\pb{\ifm \bar{p} \else $\bar{p}$\fi}
\def\xg{\ifm x_{\gam} \else $x_{\gam}$\fi}
\def\xp{\ifm x_p \else $x_p$\fi}
\def\cdh{\ifm {\cal D}_h^H \else $ {\cal D}_h^H$\fi}
\def\dh{\ifm D_h^H \else $D_h^H$\fi}
\def\ch{\ifm C_{f,h}^H \else $C_{f,h}^H$\fi}
\def\eps{\ifm \epsilon \else $\epsilon $\fi}
\def\fac{
(\frac{N}{2\pi}) \frac{1}{\Gamma (1-\epsilon )}
(\frac{4\pi \mu^2}{s_{\rm min}})^\epsilon \frac{1}{\eps }}
\def\GeV{\rm GeV}
\newcommand{\notp}{\ \hbox{{$p$}\kern-.43em\hbox{/}}}
\newcommand{\notE}{\ \hbox{{$E$}\kern-.43em\hbox{/}}}
\newcommand{\beq}{\begin{equation}}
\newcommand{\eeq}{\end{equation}}
\newcommand{\beqn}{\begin{eqnarray}}
\newcommand{\eeqn}{\end{eqnarray}}
%same but with no equation numbers
\newcommand{\beqs}{\begin{eqnarray*}}
\newcommand{\eeqs}{\end{eqnarray*}}
\newcommand{\half}{\textstyle{1\over 2}}
\newcommand{\third}{\textstyle{1\over 3}}
\newcommand{\etal}{{\rm et al.}}                   
%   ...
\newcommand{\ttbs}{\char'134}
\newcommand{\AmS}{{\protect\the\textfont2
  A\kern-.1667em\lower.5ex\hbox{M}\kern-.125emS}}

\thispagestyle{empty}
\font\twelverm=cmr12
\font\elevenit=cmti10 scaled\magstep 1
{\twelverm
\begin{flushright}
\hfill{CERN-TH/96-146}\\
\hfill{FERMILAB-conf/96-129-T\\
hep-ph/9606209}
\end{flushright}
\vskip 2cm
\begin{center}
 W PLUS HEAVY QUARK PRODUCTION AT THE TEVATRON
\vglue 1.4cm
\begin{sc}
{\twelverm Walter T. Giele, Stephane Keller}\\
\vglue 0.2cm
\end{sc}
{\elevenit Fermilab, MS 106\\
Batavia, IL 60510, USA}
\vglue 0.5cm
and
\vglue 0.5cm
\begin{sc}
{\twelverm Eric Laenen}\\
\vglue 0.2cm
\end{sc}
{\elevenit CERN TH-Division\\
1211-CH, Geneva 23, Switzerland}
\end{center}

\vglue 1cm
\par \vskip .1in \noindent
\renewcommand{\baselinestretch}{1.1}
We present the next-to-leading order QCD corrections to  
the production of a $W$-boson in association with a jet containing a heavy
quark. The calculation is fully
differential in the final state particle momenta and includes
the mass of the heavy quark.  We study for the case of the Tevatron 
the sensitivity of the cross section to the strange quark distribution
function, the dependence of the cross section on the heavy quark mass,
the transverse momentum distribution of the 
jet containing the heavy quark, and the momentum distribution
of the heavy quark in the jet.
\vglue 1cm

\center{\elevenit

To appear in:\\ Proceedings of the DESY-Zeuthen
Workshop: ``QCD and QED in Higher Orders'',
Rheinsberg, Germany, April 1996.
}

\vfill
\begin{flushleft}
CERN-TH/96-146\\
FERMILAB-conf/96-129-T\\
June 1996
\end{flushleft}
}

\newpage
\setcounter{page}{1}

% add words to TeX's hyphenation exception list
\hyphenation{author another created financial paper re-commend-ed}

% declarations for front matter

\title{W plus heavy quark production at the Tevatron
\thanks{Talk presented by E. Laenen.}}

\author{Walter T. Giele, Stephane Keller\address{Fermilab, MS 106 \\
        Batavia, IL 60510, USA}%
        and
        Eric Laenen\address{CERN TH-Division\\
        CH-1211 Geneva 23, Switzerland}}

\begin{abstract}
We present the next-to-leading order QCD corrections to  
the production of a $W$-boson in association with a jet containing a heavy
quark. The calculation is fully
differential in the final state particle momenta and includes
the mass of the heavy quark.  We study for the case of the Tevatron 
the sensitivity of the cross section to the strange quark distribution
function, the dependence of the cross section on the heavy quark mass,
the transverse momentum distribution of the 
jet containing the heavy quark, and the momentum distribution
of the heavy quark in the jet.
\end{abstract}

\maketitle

\section{INTRODUCTION}

\noindent The study of jet production in association with a vector boson
at hadron colliders has been succesful in the recent past.
The advantage of this signal over pure jet production is that 
the lepton(s) from the vector boson decay can be used
as a trigger, such that jets can be studied free from jet-trigger bias.
Furthermore, the lower rate obviates the need for prescaling.  
On the theory side, progress in calculational
techniques to construct next-to-leading order (NLO) Monte-Carlo programs~\cite{GGK93} has allowed a meaningful confrontation with data~\cite{D0W}.
For a review of such techniques and more recent progress, see \cite{KKCS}.

The tagging of heavy hadrons in the jet 
offers a unique possibility of studying the hadronic structure inside the jet.
By considering jets where the {\em leading} hadron is tagged, a clear connection
can be made with perturbative QCD.  
At the parton level, the tagging of a heavy hadron corresponds to the tagging 
of a heavy-flavor quark. Experimentally, 
the presence of a $D$ or $B$ meson is inferred through its decay products.
The importance of heavy flavor tagging
has been clearly demonstrated in the analysis that led to the top 
quark discovery \cite{CDFD0}.  In the future, heavy flavor tagging will 
continue to be an important analysis tool. It will provide more detailed 
information about the event, and test of the underlying QCD theory.

If one demands the presence of a {\em charm} quark in the
jet recoiling against a $W$-boson, the signal is directly 
sensitive to the strange quark distribution function in 
the proton, at a scale of the order of    
the $W$-mass.  

Replacing the charm quark by a {\em bottom} quark, 
one could use this process as an alternative calibration of the $b$-quark
tagging efficiency.  This will be useful at luminosities achieved by 
the Main Injector.
However, by far 
the dominant contribution to $W$+bottom production is due to $W+b\bar{b}$ 
production, where the heavy quark pair is produced by gluon splitting.
Here the inclusion of the gluon to $B$ meson 
fragmentation function is probably more important than the inclusion of 
NLO effects to the $W$+bottom process, which
implies that this reaction can constrain
this fragmentation function.

In this report, which is based on Ref.~\cite{GKL1}, 
we present the calculation of the QCD corrections up
to $O(\alpha_s^2)$ of the process $p\bar{p}\rightarrow W+Q$
where $Q$ is an heavy quark, keeping the mass of the heavy
quark explicit, and the final state fully exclusive.

\section{METHOD}

\noindent
The method we used to calculate the 
$O(\alpha_s^2)$ QCD corrections to 
$p\bar{p}\rightarrow W +Q$ 
is a generalization \cite{KL} of the phase
space slicing method of Ref.~\cite{GG92}
to include massive quarks.  

The leading order (LO) calculation involves the simple
subprocess $q+g \rightarrow W + Q$, where $q$ is a light quark
and $g$ a gluon, at tree level.

The virtual corrections consist of 
the interference between the lowest order diagrams and their one-loop 
corrections. The integrals over the loop momenta
were performed in $d=4-2\epsilon$ dimensions.
Ultraviolet singularities were absorbed through mass (on-shell scheme)
and coupling-constant (modified $\overline{\rm MS}$ scheme \cite{CWZ})
renormalization.  
The remaining soft and collinear singularities,
appearing as $1/\eps^2$ and $1/\eps$ poles, factorize into a universal 
factor multiplying the Born cross section.  

The real corrections
consist of the contributions from all the subprocesses $i\,j \rightarrow W\,Q\,k$,
where $i,j,k$ are massless partons, and the subprocess 
$i\,j \rightarrow W\,Q\,\bar{Q}$~\cite{Mang}.
Some of these contributions exhibit soft and/or collinear singularities.  
We do not consider diagrams where the heavy quark is in the initial state.
The method we used to isolate the singularities consists of 
slicing up the phase space, using a cut-off $s_{\rm min}$
and color-ordered subamplitudes,
into a hard region ($s_{\rm min} < 2p_r\cdot p_s$ for any $r,s$), 
containing no singularities, and a complementary 
region in which the final state parton
is either soft or emitted collinearly with one of the initial state partons.  
In the hard phase space region, one 
can work in four dimensions and perform the phase space 
integration numerically.  
In the soft and collinear region, the integration is done analytically in $d$
dimensions using soft and collinear approximations, which are valid in the 
limit that $s_{\rm min}$ is small.  The cross section in this region again 
factorizes into a universal factor multiplying the Born cross section.  
The initial state collinear singularities are factorized 
into parton distribution functions
in the ${\rm\overline{MS}}$ scheme, using the formalism
of crossing functions~\cite{GGK93}.

Note that the process $i\,j \rightarrow W\,Q\,\bar{Q}$ is quite   
different from the other subprocesses: the heavy quark does not 
originate from the $W$ vertex, and it is independent of 
$s_{\rm min}$, because it is free from singularities.

Adding the real and virtual corrections leads to the cancellation
of all remaining singularities.  We checked gauge invariance for both the 
virtual and real corrections.
Finally, we constructed a Monte Carlo program for the present
process, including these corrections.

Before showing any numerical results,
we first list here the default choices we made for parameters and cuts in
producing the results of this paper.  Any deviation from these
choices will be indicated explicitly.
For the case of charm (bottom) we assumed three (four) light flavors.
We used both at LO and NLO the CTEQ3M~\cite{CTEQ3M} set of parton distribution functions, 
and a two-loop running coupling constant with
$\Lambda^{(2,4)}_{\rm QCD}$ the
value supplied with the CTEQ3M set ($0.239$ GeV).
We implemented continuity across heavy flavor thresholds \cite{CT86} 
using the parametrization of Ref.~\cite{ADMN}.
We used the Snowmass convention~\cite{SNOW} for the definition of a jet. 
Our conditions on the transverse energy and pseudorapidity of the
jet were $E_{T}({\rm jet}) > 10$ GeV, $|\eta_{jet}| < 3$,
and no cuts on the W.
We took a jet cone size of $\Delta R=0.7$,
the $W$ mass equal to $m_W=80.23$~GeV, and
the heavy quark mass $m$ equal to $1.7$~GeV for charm 
and $5$~GeV for bottom. We used $V_{cs}=0.97$
and  $V_{cd}=0.22$ for the relevant Cabibbo-Kobayashi-Maskawa matrix elements.
We chose the factorization scale equal to the renormalization scale 
and denote it by $\mu$, taking $\mu=m_W$.
At least one heavy quark was required to be inside of the jet, with the sign
of its electric charge correlated with the $W$ charge, as in the LO case.

We verified the independence of the calculation on
the arbitrary parameter $s_{\rm min}$.
The results given in the remainder of this report are
averaged over $s_{\rm min}$ between 1 and 10 GeV. 

We further found the scale dependence of the inclusive cross section
to be somewhat reduced by the inclusion of the NLO corrections.

\section{STRANGE QUARK DISTRIBUTION FUNCTION}

\noindent
We first discuss the effect that the inclusion of the NLO QCD corrections
to $W$ + charm-tagged jet production has on constraining the strange quark distribution
function $s(x,\mu)$ in the proton.  Here $x$ is the momentum fraction of the 
strange quark in the proton, and $\mu$ is the factorization scale.
See Ref.~\cite{BHKMR} for an extensive study of this issue 
using the shower Monte-Carlo program PYTHIA \cite{PYT87}.
\begin{table*}[hbtp]
% space before first and after last column: 1.5pc
% space between columns: 3.0pc (twice the above)
\setlength{\tabcolsep}{1.5pc}
% -----------------------------------------------------
% adapted from TeX book, p. 241
\newlength{\digitwidth} \settowidth{\digitwidth}{\rm 0}
\catcode`?=\active \def?{\kern\digitwidth}
% -----------------------------------------------------
\caption{The $W$ + charm-tagged one-jet inclusive cross section in pb for
LO, $W+Q\bar{Q}$, and NLO (including the $W+Q\bar{Q}$ contribution)
using  different sets of
parton distribution functions.  The statistical uncertainty
from the Monte-Carlo integration is less than 1\%.}
\begin{tabular*}{\textwidth}{@{}l@{\extracolsep{\fill}}ccccc}
\hline 
  set           &mass (GeV)    & LO    & $WQ\bar{Q} $  &  NLO \\ 
  CTEQ1M        &$m_c$=1.7      & 96    &20             &161\\ 
  MRSD0'        &$m_c$=1.7      & 81    &20             &138\\ 
  CTEQ3M        &$m_c$=1.7      & 83    &20             &141\\ 
  CTEQ3M        & $m_b$=5.0     & 0.17   &9.09           &9.33 \\ 
\end{tabular*}
\end{table*}

In Table 1 we give the NLO cross section for the parton distribution function sets
CTEQ1M \cite{CTEQ1M} and MRSD0' \cite{MRSD0}.  
The MRSD0' set derives its strange quark distribution 
from di-muon data, whereas the CTEQ1M set uses DIS data.
At low $\mu^2$ and low $x$ the difference is as much
as a factor of two.
Also shown is the result obtained with the more recent CTEQ3M set, which uses
the same assumption about the strange quark distribution as MRSD0'.    
Comparing the CTEQ3M and MRSD0' sets, we see that the difference due to using more
recent data sets in the global fit for the parton distribution functions is small.
This is also reflected in the cross section for $W+ c\bar{c}$, which is 
the same for all three sets.  We can conclude that the difference between 
CTEQ1M and MRSD0' of 15.4 \% is due to the strange quark 
distribution function.
This difference becomes 14.5\%  when one includes the 
$W+b$ background (9 pb, almost all of it
coming from the gluon splitting contribution, see Table 1), and
assuming conservatively that each bottom quark 
is mistagged as a charm quark.
This shows that the conclusions reached in Ref.~\cite{BHKMR}
are still valid at NLO. In both the NLO calculation and the PYTHIA analysis
about 50\% of the contributions are initiated by strange quarks.
One major difference is that PYTHIA suggests that the gluon splitting 
contributes about 35\%, 
whereas in the NLO calculation it is only about 15\%.  
This number is however quite 
sensitive to the choice of scale $\mu$
in the gluon splitting contribution.  

\section{MASS DEPENDENCE OF JET PRODUCTION}

\noindent
Next we compare $W$ + untagged jet production\footnote{
For the untagged process we take five massless quark flavors.}  \cite{GGK93}
and $W$ + charm-tagged 
jet production. In particular, we study
the jet transverse energy ($E_T({\rm jet})$) distribution.
In Fig.~\ref{fig:ratio} we show the ratio of the charm-tagged jet 
over the untagged jet $E_T({\rm jet})$-distribution for the LO and NLO cases. 
At LO the charm-tagged jet is simply represented by a charm quark.
The ratio in Fig.~\ref{fig:ratio} has a characteristic shape which 
can readily be understood at tree level.
\begin{figure}[hbtp]
\vspace{4cm}
\begin{picture}(7,7)
\includegraphics{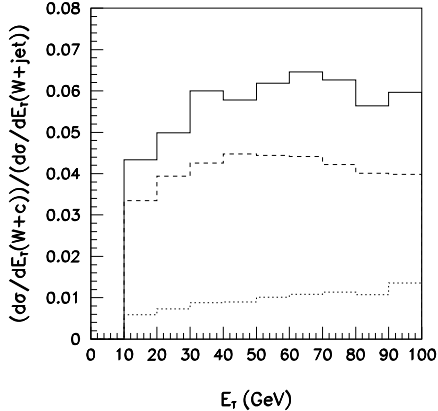}
\end{picture}
\caption{
Ratio of $W$ + charm-tagged inclusive one-jet 
production over $W$ + inclusive untagged one-jet production 
as function of the jet transverse energy.
The solid line is the NLO ratio and the dashed line the LO ratio.  
The $W+c\bar{c}$ contribution to the NLO ratio is also shown 
(dotted line).  
}
\label{fig:ratio}
\end{figure}
At low $E_T({\rm jet})$ the charm-tagged jet rate is suppressed relative to
the untagged-jet rate due to its fermionic final state. The untagged-jet 
rate is dominated by the gluonic final state which has a soft singularity,
that is absent for the fermionic final state.
At high $E_T({\rm jet})$ we again observe a relative suppression of the 
charm-tagged jet because at LO this process has a gluon in the
initial state. At high $E_T({\rm jet})$ 
the dominant scattering in the untagged-jet rate 
is due to quark-antiquark 
collisions, again favoring the gluonic final state. 
Clearly, apart from an approximate overall $K$-factor, the NLO cross section retains these features.
Also shown in Fig.~\ref{fig:ratio} is the $W$ + $c\bar c$ contribution. 
At low $E_T({\rm jet})$ 
its suppression is more pronounced
due to the charm quark pair production threshold. 
At high $E_T({\rm jet})$ there is no suppression 
for this process because it is instigated by a quark-antiquark 
collision.

We now turn to the mass dependence of the cross section.
We will show that there are important mass effects 
at NLO, especially when we look in detail at the tagged jets.
(At LO one may safely take the mass to zero; the answer
does not change.)

Let us first consider the $W+Q\bar{Q}$ contribution.    
Because the mass of the heavy quark regulates the collinear singularity, 
it is expected that the strongest mass dependence will come from the collinear 
region.
In this region the cross section factorizes
into the cross section
for $W + {\rm gluon}$ production  multiplied by a universal factor.  
After integration over the invariant mass of the
heavy quark pair we find
that the mass dependent part of this universal factor has the following form:
\beq
\als \frac{N_c}{8\pi} P_{q\overline{q} \go g} (z) 
\ln\left( \frac{M^2}{m^2} \right)  
dz
\label{eq:lnm2}
\eeq 
where $M$ is the upper limit of the 
heavy quark pair invariant mass defining the collinear region,  
and $P_{q\overline{q} \go g} (z)$ is the massless 
Altarelli-Parisi~\cite{AP} splitting function:
\beq
P_{q\overline{q} \go g} (z) = \frac{2}{N_c} (z^2+(1-z)^2).
\label{eq:p}
\eeq 
We chose the following definition of $z$
\beq
z=\frac{E+P_{\parallel}}{E_{jet}+P_{jet}}
\eeq
where $E$ and $P_{\parallel}$ are the heavy quark energy and momentum 
projected on the jet direction, 
and $E_{jet}$ and $P_{jet}$ are the jet energy and momentum.
Other choices, such as $z=E_T(Q)/E_T({\rm jet})$,
do not change any of the conclusions in what follows.
In the strictly collinear limit $M$ is much 
smaller than the energy of the gluon, 
but in a leading logarithmic 
\begin{figure}[hbtp]
\vspace{3.5cm}
\begin{picture}(7,7)
\includegraphics{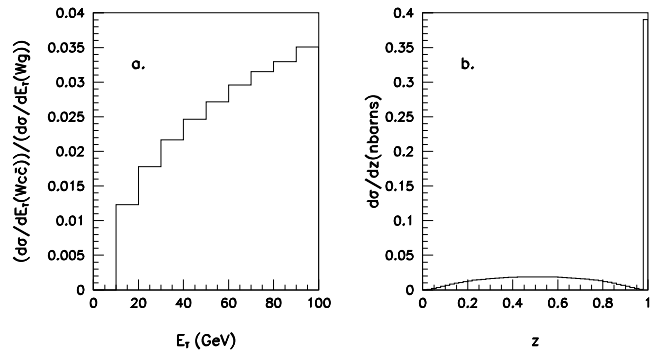}
\end{picture}
\caption{
a) Ratio of the $W+c \bar{c}$ component of the charm-tagged one-jet inclusive
cross section to the
$W+gluon$ cross section, as a function of the 
jet tranverse energy.  b) The $z$-distribution of the 
$W+c \bar{c}$ component.
}
\label{fig:wcc}
\end{figure}
approximation one may take $M$ to be of the order of $E_T({\rm jet})$.
The behavior of Eq.~(\ref{eq:lnm2}) can be seen 
qualitatively in Fig.~\ref{fig:wcc}a, where the 
ratio of the $W+c\bar{c}$ cross section over the $W+{\rm gluon}$ cross section is 
shown as a function of the transverse energy of the jet.  
One can see an approximate logarithmic enhancement with increasing $E_T$,
as predicted by  the leading logarithmic approximation.
However the $z$-distribution, plotted in Fig.~\ref{fig:wcc}b, 
does not conform with
the $z$-dependence in Eq.~(\ref{eq:lnm2}).    
First, the peak at $z=1$ is due to events where the $\bar{Q}$ is not inside 
of the jet, such that the whole jet is formed by the lone $Q$.  
Second, the cross section is suppressed near $z=0$ and $z=1$ (excluding the 
peak) due to terms in the collinear region that depend strongly on $z$,
but not on $M$.
One way to enhance the effect of the 
leading logarithm term of Eq.~(\ref{eq:lnm2}) 
in the $z$ distribution is to lower the mass of the heavy quark in our calculation.  
This is done in Fig.~\ref{fig:z}a, where we show the $z$-distribution 
for the case $m=0.01$ GeV. Clearly it now resembles the functional form
of Eq.~(\ref{eq:p}) much closer.

The $\ln(m^2)$ term in Eq.~(\ref{eq:lnm2}) diverges in the limit of vanishing 
quark mass, and is not cancelled by any other contribution~\footnote{
In the $W$ +1 jet calculation, this singularity  
is cancelled by a companion collinear singularity in the quark-loop correction
to the outgoing gluon in the Born diagram. In our calculation this diagram is
not present.}.
In principle, any observable should be ``collinear safe'', i.e.
if the mass is taken to zero, the observable should 
be finite and approach the massless result.  
This is needed to describe situations where the relevant scale
is much larger than the heavy quark mass. 
In the present case we are dealing with a final state divergence in $\ln(m^2)$
that should be factorized 
into the fragmentation function of a gluon into a heavy hadron.  
The evolution of fragmentation functions will then
resum the large logarithms $\ln(E_T^2)$.
This problem is the final state version of the problem of heavy quark 
distribution functions~\cite{ACOT94}.  It 
will be discussed in more detail in \cite{KL}.
Some studies in this regard were done in Ref.~\cite{Greco}.
Here, we simply keep the mass finite.
\begin{figure}[hbtp]
\vspace{3.3cm}
\begin{picture}(7,7)
\includegraphics{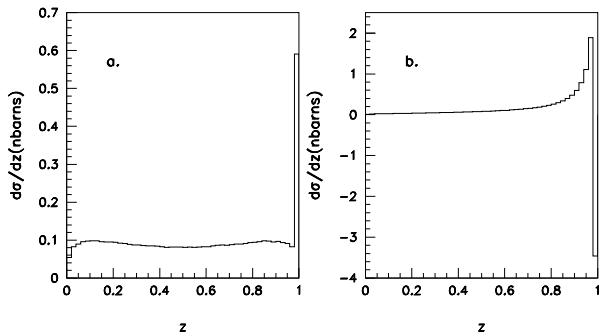}
\end{picture}
\caption{
The $z$-distribution of the charm-tagged one-jet inclusive cross section,
with $m=0.01$ GeV.  a) $Wc\bar{c}$ component. b) Total 
contribution minus the $Wc\bar{c}$ component.
}
\label{fig:z}
\end{figure}
Let us now turn to the NLO single charm contribution (excluding the 
$W+c\bar{c}$ contribution).  In the collinear region, the cross section again
factorizes, leading, after integration over
the invariant mass of the collinear partons, to the universal factor:
\beq
\als \frac{N_c}{8\pi} P_{qg \go q} (z) 
\ln\left( \frac{M^2}{m^2} \right)  
\label{eq:scll}
dz
\eeq  
where $P_{qg \go q} (z)$ is the splitting function:
\beqn
P_{qg \go q} (z) = \lim_{\delta\rightarrow 0}2 (1-\frac{1}{N_c^2}) 
\Big(\left(\frac{1+z^2}{1-z}\right)\times\nonumber\\ \theta(1-z-\delta)
+(\frac{3}{2}+2\ln\delta)\,\delta(1-z)\Big).
\label{eq:pqg}
\eeqn
In Fig.~6b we show for the single charm quark contribution the
$z$-distribution, with the contribution of Eq.~(\ref{eq:scll}) enhanced
by taking  $m = 0.01$ GeV. Note that it resembles the functional
form in Eq.~(\ref{eq:pqg}).  From Eq.~(\ref{eq:pqg}) one derives
\beq
\int_0^1 P_{qg \go q} (z) dz= 0 \ ,
\label{eq:qsum}
\eeq  
which must hold for the probability to find a quark in a quark of the same 
flavor to be one \cite{AP}.  
From Eqs.~(\ref{eq:qsum}) and (\ref{eq:scll}) 
we can now make the important observation that as long as the cuts 
on the heavy quark-tagged jet are such that all $z$-values are allowed to 
contribute, 
there are no large logarithms $\ln(E_T^2/m^2)$.  
\begin{figure}[hbtp]
\vspace{3.5cm}
\begin{picture}(7,7)
\includegraphics{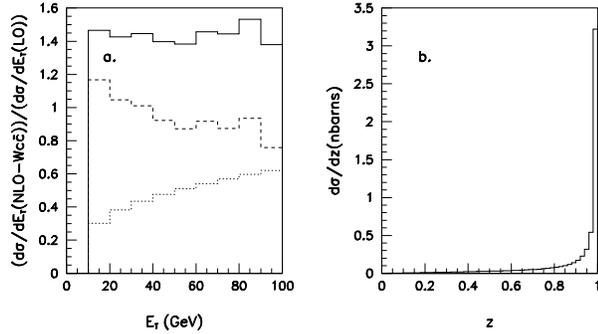}
\end{picture}
\caption{
a) $E_T$ distribution ratio of NLO charm-tagged one-jet inclusive 
cross section to LO one.
Solid line has no restrictions on the $z$-integration, 
the dotted and dashed line have restrictions of $z<0.9$ and $z>0.9$ 
respectively.
b) $z$-distribution of NLO charm-tagged one-jet inclusive cross section.  
The $W+c\bar{c}$ component is not included in these plots.
}
\label{fig:z+et}
\end{figure}
However, if the cuts are such that the $z$-integration is restricted, or 
convoluted with a $z$-dependent function, some 
$\ln(E_T^2/m^2)$ terms will remain.  An example of such a 
convolution is the $E_T$ distribution of the heavy quark itself.    
All this is illustrated in Fig.~\ref{fig:z+et}a, 
where two cases of $z$-restrictions ($z > 0.9$ and $z < 0.9$) are plotted
in addition to the all-$z$ case. 
Note that for the all-$z$ case
the ratio is indeed essentially independent of $E_T({\rm jet})$, but that for 
the $z$-restricted cases the logarithmic dependence is apparent. 
For completeness we show in Fig.~\ref{fig:z+et}b the $z$-distribution
for the single charm contribution for $m=1.7$ GeV.

Thus care must be taken when calculating tagged cross sections 
in determining whether or not there are large logarithms present 
due to restrictions on the $z$-integration. 
Such restrictions would also necessitate the introduction of the appropriate 
fragmentation function to absorb 
the $\ln(m^2)$ terms, as in the $WQ\bar{Q}$ case.

Note that for the example of the jet transverse energy distribution
given in Fig.~4 there is no constraint on the $z$-integration, so that the
only final state $\ln(m^2)$ term arises from the $Wc\bar{c}$ contribution.

\section{CONCLUSIONS}

\noindent
We have presented the calculation 
of the $O(\alpha_s)$  corrections to the reaction
$p\bar{p}\rightarrow W +Q$.  
We found that the inclusion of the NLO corrections
does not change the conclusions of Ref.~\cite{BHKMR} about constraining 
the strange quark distribution function using
$W$ + charm-tagged jet events at the Tevatron.
However, using our NLO calculation, this procedure can
constrain the NLO strange quark distribution function, 
once a reasonable data sample is collected.
We studied the $E_T$ distribution
of the jet containing the heavy quark and the mass dependence of the cross 
section, and noted the need to include the 
heavy hadron fragmentation functions.

\end{document}